\documentstyle[12pt]{article}

\def\thefootnote{\fnsymbol{footnote}}

\newlength{\minitwocolumn}
\catcode`\@=11
\long\def\@makefntext#1{
\protect\noindent \hbox to 3.2pt {\hskip-.9pt  
$^{{\eightrm\@thefnmark}}$\hfil}#1\hfill}               

\def\thefootnote{\fnsymbol{footnote}}
\def\@makefnmark{\hbox to 0pt{$^{\@thefnmark}$\hss}}    
        
\def\ps@myheadings{\let\@mkboth\@gobbletwo
\def\@oddhead{\hbox{}
\rightmark\hfil\eightrm\thepage}   
\def\@oddfoot{}\def\@evenhead{\eightrm\thepage\hfil
\leftmark\hbox{}}\def\@evenfoot{}
\def\sectionmark##1{}\def\subsectionmark##1{}}

\font\eightrm=cmr8

\def\MPL{Mod.~Phys.~Lett. }

\def\PL{Phys.~Lett. }
\def\PR{Phys.~Rev. }

\def\PTP{Prog.~Theor.~Phys. }

\def\brs{\delta}

\newcommand{\bracket}[2]{\langle #1\,,#2\rangle}

\newcommand{\calA}{{\cal A}}
\newcommand{\calB}{{\cal B}}
\newcommand{\calC}{{\cal C}}
\newcommand{\calD}{{\cal D}}

\newcommand{\calI}{{\cal I}}
\newcommand{\calJ}{{\cal J}}
\newcommand{\calL}{{\cal L}}

\newcommand{\Q}{{\kern.24em\vrule width.04em height1.4ex%
                 depth-.05ex\kern-.26em\mathsf Q}}
\newcommand{\C}{{\kern.24em\vrule width.04em height1.4ex%
                 depth-.05ex\kern-.26em\mathsf C}}

\begin{document}


\baselineskip 0.7cm

\begin{titlepage}
\begin{flushright}
\end{flushright}

\vskip 1.35cm
\begin{center}
{\Large \bf
Three Dimensional Topological Field Theory
induced from Generalized Complex Structure}
\vskip 1.2cm
Noriaki IKEDA$^1$%
\footnote{E-mail address:\ ikeda@yukawa.kyoto-u.ac.jp}
\vskip 0.4cm
{\it $^1$Department of Mathematical Sciences, 
Ritsumeikan University \\
Kusatsu, Shiga 525-8577, Japan }\\

\date{}

\vskip 1.5cm

\begin{abstract}
We construct a three-dimensional topological sigma model which is 
induced from a generalized complex structure on a target generalized 
complex manifold.
This model is constructed from maps from a 
three-dimensional manifold $X$ 
to an arbitrary generalized complex manifold $M$.
The theory is invariant under the diffeomorphism on the world volume
and the $b$-transformation on the generalized complex structure.
Moreover the model is manifestly invariant under the mirror symmetry.

We derive from this model
 the Zucchini's two dimensional topological sigma model with 
a generalized complex structure as a boundary action 
on $\partial X$.
As a special case, we obtain three dimensional realization of a 
WZ-Poisson manifold.
\end{abstract}
\end{center}
\end{titlepage}

\renewcommand{\thefootnote}{\alph{footnote}}

\setcounter{page}{2}


\rm
\section{Introduction}
\noindent
A generalized complex structure is introduced by Hitchin \cite{Hit}
to unify the complex and symplectic geometry, and studied in detail
by Gualtieri in \cite{Gua}.

It is found that 
Geometry of the $N = (2, 2)$ supersymmetric sigma model \cite{GHR} 
has been formulated by a generalized K\"ahler geometry, 
which is a generalization of K\"ahler geometry to 
a generalized complex geometry.
Noncommutative deformation of 
$N = 2$ supersymmetric sigma model on a Calabi-Yau manifold 
and its topological twisted sigma model can be described in terms of 
a generalized complex structure \cite{Kap}\cite{KL}.
\cite{Lin}\cite{LMTZ}\cite{Zab}\cite{Lindstrom:2004hi}
have studied a realization of a generalized complex structure 
by a $N = (2,2)$ supersymmetric sigma model 
and a topological sigma model.
In \cite{FMT}\cite{GMPT}\cite{Grange:2004ah}, it has been investigated that 
the supersymmetric $SU(3)$ manifolds 
are generalized Calabi-Yau manifolds.
Since  mirror symmetry is correspondence of complex 
and symplectic manifolds,
the generalized complex geometry is considered as 
a natural framework of mirror symmetry.
Mirror symmetry on a generalized complex geometry is investigated in
\cite{Bas}\cite{Jes}\cite{CGJ}.
The relation of the generalized complex structure with the supersymmetric 
Poisson sigma model has been discussed in \cite{Bergamin:2004sk}.

In this paper, we propose a realization of a generalized 
complex structure by a three dimensional topological 
sigma model (a topological membrane).

We consider that a three dimensional topological field theory is 
a natural framework to analyze a generalized complex structure.
Since mirror symmetry exchange $A$ model and $B$ model,
it seems to be natural that the model with a generalized complex structure
has $M$-theoretic, or membrane theoretic realization.
Integrability condition of a generalized complex structure is defined by a 
Courant bracket \cite{Courant}. On the other hand,
three dimensional topological sigma model naturally 
has a Courant algebroid structure 
\cite{Ikeda:2002qx}\cite{Hofman:2002rv}.
We will find that 
a $3$-form $H$ in a twisted generalized complex structure is naturally 
introduced in three dimensional topological sigma model.

Recently, Zucchini has constructed a two-dimensional 
topological sigma model induced from a target generalized 
complex structure as a generalization of a Poisson sigma model
\cite{Zucchini:2004ta}.
However a generalized complex structure is a sufficient but 
not necessary condition in his model.
Our model solves this nonequivalence.
Moreover we consider a direct relation of our model with 
the Zucchini's model.
If we consider a three dimensional world volume $X$ with a two dimensional 
boundary $\Sigma = \partial X$, 
we can derive the Zucchini action as a boundary action 
on $\Sigma$ of our model on $X$.
As a special case,
we can realize a WZ-Poisson sigma model
\cite{KS} as 
a boundary action of a three dimensional topological sigma model.

The paper is organized as follows. 
In section 2, we summarize about a generalized complex structure.
In section 3, we define a model induced from a generalized complex structure.
We propose an action of a three dimensional topological sigma model with 
a generalized complex structure.
We also consider the action with a twisted generalized complex structure.
In section 4, we analyze the relation with the 3D nonlinear gauge theory.
In section 5, we derive the Zucchini's two dimensional model 
with a generalized complex structure as a boundary action of our model.
Section 6 is conclusion and discussion.

\section{Generalized Complex Structure}
\noindent
In this section, we summarize a generalized complex structure, 
based on description of section 3 in \cite{LMTZ} 
and section 2 in \cite{Zucchini:2004ta}.

Let $M$ be a manifold of even dimension $d$ with 
a local coordinate $\{ \phi^i \}$. We consider the vector bundle 
$TM \oplus T^*M$. 
We consider a section $X + \xi \in C^\infty(TM\oplus T^*M)$ 
where $X \in C^\infty(TM)$ and $\xi \in C^\infty(T^*M)$.

$TM \oplus T^* M$ is equipped with a natural indefinite metric of signature  
$(d,d)$ defined by 
\begin{eqnarray}
\langle X + \xi , Y+\eta \rangle= \frac{1}{2} (i_X \eta + i_Y \xi),
\end{eqnarray}
for $X+\xi, Y+\eta \in C^\infty (TM \oplus T^* M)$, where $i_V$ 
is an interior product with a vector field $V$. 
In the Cartesian coordinate $(\partial/\partial \phi^i, d \phi^i)$, 
we can write the metric as follows
\begin{eqnarray}
{\calI} = \left(\matrix{0 & 1_d &\cr
                          1_d & 0 &\cr}\!\!\!\!\!\!\!\right),
\end{eqnarray}

We define a Courant bracket on $TM \oplus T^* M$,
\begin{eqnarray}
[ X + \xi, Y + \eta] = [X,Y] + \calL_X \eta - \calL_Y \xi 
- \frac{1}{2} d_M (i_X\eta-i_Y\xi),
\end{eqnarray}
with $X+\xi, Y+\eta\in C^\infty(TM\oplus T^* M)$, where $\calL_V$ 
denotes Lie derivation with respect a vector field $V$ and $d_M$ is 
the exterior differential of $M$.
This bracket is antisymmetric but do not satisfy the Jacobi identity.
We may define a so called Dorfman bracket as follows
\begin{eqnarray}
(X+\xi) \circ (Y+\eta) = [X,Y]+ \calL_X \eta - i_Y d \xi,
\end{eqnarray}
which satisfies the Jacobi identity but is not antisymmetric.
Antisymmetrization of a Dorfman bracket coincides with a Courant bracket.


A generalized almost complex structure $\calJ$ is a section of 
$C^\infty ({\rm End}(TM\oplus T^* M))$, which is an isometry of 
the metric $\langle\,,\rangle$, 
$
{\calJ}^* \calI \calJ = \calI,
$
and satisfies 
\begin{eqnarray}
{\calJ}^2 = -1.
\end{eqnarray}

A $b$-transformation is an isometry defined by 
\begin{eqnarray}
\exp(b)( X + \xi)= X + \xi + i_X b,
\end{eqnarray}
where $b \in C^\infty(\wedge^2 T^* M)$ is a $2$--form. 
A Courant bracket is covariant under the $b$-transformation
\begin{eqnarray}
[\exp(b) (X + \xi), \exp(b)(Y + \eta)] = \exp(b)[X + \xi, Y + \eta],
\end{eqnarray}
if the $2$--form $b$ is closed.
%
The $b$-transform of $\calJ$ is defined by 
\begin{eqnarray}
\hat{\calJ}=\exp(-b) {\calJ} \exp(b).
\end{eqnarray}

$\calJ$ has the $\pm \sqrt{-1}$ eigenbundles.
Therefore we need complexification of $TM \oplus T^* M$, 
$(TM \oplus T^* M) \otimes \C$. 
The projectors on the eigenbundles are constructed by 
\begin{eqnarray}
\Pi_{\pm} = \hbox{$1\over 2$}(1\mp\sqrt{-1}{\calJ}).
\end{eqnarray}
The generalized almost complex structure $\calJ$ is integrable if 
\begin{eqnarray}
\Pi_{\mp}[\Pi_{\pm}(X+\xi),\Pi_{\pm}(Y+\eta)]=0,
\label{integrability}
\end{eqnarray}
for any $(X+\xi),(Y+\eta) \in C^\infty(TM \oplus T^* M)$,
where the bracket is the Courant bracket.
Then $\calJ$ is called a generalized complex structure. 
Integrability is equivalent to the single statement
\begin{eqnarray}
N(X+\xi,Y+\eta)=0,
\end{eqnarray}
for all $X+\xi, Y+\eta\in C^\infty(TM\oplus T^* M)$, 
where $N$ is the generalized Nijenhuis tensor defined by 
\begin{eqnarray}
N(X+\xi,Y+\eta)
&=& [X+\xi,Y+\eta] -[{\calJ}(X+\xi), {\calJ}(Y+\eta)]
+{\calJ}[{\calJ}(X+\xi),Y+\eta]
\nonumber \\
&&\, +{\calJ}[X+\xi, {\calJ}(Y+\eta)].
\end{eqnarray}
The $b$-transform $\hat{\calJ}$ of a generalized complex structure $\calJ$ is 
a generalized complex structure if the $2$--form $b$ is closed. 

We decompose a generalized almost complex structure 
$\calJ$ in coordinate form as follows 
\begin{eqnarray}
{\calJ} = \left(\matrix{J& P&\cr
                          Q& -J^*&\cr}\!\!\!\!\!\!\!\right),
\label{gcstcomp}
\end{eqnarray}
where $J\in C^\infty(TM\otimes T^* M)$, $P\in C^\infty(\wedge^2 TM)$, 
$Q\in C^\infty(\wedge^2 T^* M)$.

Then the condition ${\calJ}^2 = -1$ is as follows:
\begin{eqnarray}
&& 
J^i{}_k J^k{}_j + P^{ik}Q_{kj} + \delta^i{}_j = 0,
\nonumber \\
&& 
J^i{}_k P^{kj} + J^j{}_k P^{ki} = 0,
\nonumber \\
&& 
Q_{ik} J^k{}_j + Q_{jk} J^k{}_i = 0,
\label{J2zero}
\end{eqnarray}
where 
\begin{eqnarray}
&& 
P^{ij}+P^{ji}=0,
\nonumber \\
&& 
Q_{ij}+Q_{ji}=0.
\end{eqnarray}
The integrability condition (\ref{integrability}) is equivalent to the 
following condition 
\begin{eqnarray}
\calA^{ijk}= \calB_i{}^{jk} = \calC_{ij}{}^k = \calD_{ijk} = 0,
\label{intcoord}
\end{eqnarray}
where
\begin{eqnarray}
\calA^{ijk} &=& P^{il} \partial_l P^{jk} 
+ P^{jl} \partial_l P^{ki} + P^{kl} \partial_l P^{ij},
\nonumber \\
\calB_i{}^{jk} &=& J^l{}_i \partial_l P^{jk}
+ P^{jl} (\partial_i J^k{}_l - \partial_l J^k{}_i)
+ P^{kl} \partial_l J^j{}_i - \partial_i J^j{}_l P^{lk},
\nonumber \\
\calC_{ij}{}^k &=& 
J^l{}_i \partial_l J^k{}_j - J^l{}_j \partial_l J^k{}_i
- J^k{}_l \partial_i J^l{}_j + J^k{}_l \partial_j J^l{}_i
\nonumber \\
&&
+ P^{kl} (\partial_l Q_{ij} + \partial_i Q_{jl} + \partial_j Q_{li}),
\nonumber \\
\calD_{ijk} &=& J^l{}_{i} (\partial_l Q_{jk} + \partial_k Q_{lj})
+ J^l{}_j (\partial_l Q_{ki} + \partial_i Q_{lk})
\nonumber \\
&&
+ J^l{}_k (\partial_l Q_{ij} + \partial_j Q_{li})
- Q_{jl} \partial_i J^l{}_k - Q_{kl} \partial_j J^l{}_i 
- Q_{il} \partial_k J^l{}_j.
\end{eqnarray}
Here $\partial_i$ is a differentiation with respect to $\phi^i$.
The $b$--transform is 
\begin{eqnarray}
\hat J^i{}_j &=& J^i{}_j - P^{ik}b_{kj},
\nonumber \\
\hat P^{ij} &=& P^{ij},
\nonumber \\
\hat Q_{ij} &=& Q_{ij} + b_{ik} J^k{}_j - b_{jk} J^k{}_i 
+ P^{kl} b_{ki} b_{lj}.
\label{btrans}
\end{eqnarray}
where $b_{ij} + b_{ji} =0$.

The usual complex structures $J$ is embedded in 
generalized complex structures 
as the special form 
\begin{eqnarray}
 {\calJ} = \left (\matrix{J & 0 &\cr
                           0 & -J^*&\cr}\!\!\!\!\!\!\right).
\end{eqnarray}
Indeed, one can check this form satisfies conditions, 
(\ref{J2zero}) and (\ref{intcoord}) 
if $J$ is a complex structure.
Similarly, the usual symplectic structures $Q$ is obtained as 
the special form of generalized complex structures 
\begin{eqnarray}
{\calJ} = \left (\matrix{0 & -Q^{-1} &\cr
                          Q & 0&\cr}\!\!\!\!\!\!\right).
\end{eqnarray}
This satisfies (\ref{J2zero}) and (\ref{intcoord}) 
when $Q$ is a symplectic structure, i.~e.~it is closed. 
Other exotic examples exist. 
There exists manifolds which cannot support any complex or 
symplectic structure, 
but admit generalized complex structures. 

The Courant bracket on $TM \oplus T^*M$ can be modified by a closed 
$3$--form.
Let $H \in C^\infty(\wedge^3 T^*M)$ be a closed $3$--form.
We define the $H$ twisted Courant brackets by
\begin{eqnarray}
[X + \xi, Y + \eta]_H = [X + \xi , Y + \eta] + i_X i_Y H,
\end{eqnarray}
where $X+\xi, Y+\eta \in C^\infty(TM\oplus T^* M)$.
Under the $b$-transform with $b$ a closed $2$--form, 
\begin{eqnarray}
[\exp(b) (X + \xi), \exp(b)(Y + \eta)] = \exp(b)[X + \xi, Y + \eta],
\end{eqnarray}
holds with the brackets $[\,,]$ replaced by $[\,,]_H$. 
For a non closed $b$, one has
\begin{eqnarray}
[\exp(b)(X+\xi),\exp(b)(Y+\eta)]_{H - d_Mb}=\exp(b)[X+\xi,Y+\eta]_H.
\end{eqnarray}
So, the $b$-transformation shifts $H$ by the exact $3$--form $d_M b$:
\begin{eqnarray}
\hat H=H - d_M b.
\label{btransj}
\end{eqnarray}
%
%
One can define an $H$ twisted generalized Nijenhuis tensor $N_H$ as follows
\begin{eqnarray}
N(X+\xi,Y+\eta)
&=& [X+\xi,Y+\eta]_H -[{\calJ}(X+\xi), {\calJ}(Y+\eta)]_H
+{\calJ}[{\calJ}(X+\xi),Y+\eta]_H
\nonumber \\
&&\, +{\calJ}[X+\xi, {\calJ}(Y+\eta)]_H,
\end{eqnarray}
by using the brackets $[\,,]_H$ instead of $[\,,]$.
A generalized almost complex structure $\cal J$ is $H$ integrable if
\begin{eqnarray}
N_H(X+\xi,Y+\eta)=0,
\end{eqnarray}
for all $X+\xi,Y+\eta \in C^\infty(TM\oplus T^* M)$.
Then we call $\cal J$ an $H$ twisted generalized complex structure. 

The $H$ integrability conditions is obtained as
\begin{eqnarray}
\calA_H{}^{ijk} = \calB_H{}_i{}^{jk} = \calC_H{}_{ij}{}^k =
\calD_H{}_{ijk} = 0,
\end{eqnarray}
where
\begin{eqnarray}
&& 
\calA_H{}^{ijk} = \calA^{ijk},
\nonumber \\
&& 
\calB_H{}_i{}^{jk}
= \calB_i{}^{jk} + P^{jl}P^{km}H_{ilm}
\nonumber \\
&& 
\calC_H{}_{ij}{}^k 
= \calC_{ij}{}^k - J^l{}_i P^{km} H_{jlm} + J^l{}_j P^{km} H_{ilm},
\nonumber \\
&& 
\calD_H{}_{ijk} = \calD_{ijk} - H_{ijk} 
+ J^l{}_i J^m{}_j H_{klm} + J^l{}_j J^m{}_k H_{ilm}
+ J^l{}_k J^m{}_i H_{jlm}.
\end{eqnarray}


\section{Action and Symmetry}
\noindent
Let $X$ be a three-dimensional manifold
and $M$ be a target manifold
of a smooth map $\phi:X \to M$  with local
coordinate expression $\{\phi^i\}$.
We also have a vector bundle $TM \oplus T^*M$ over $M$.
We introduce $A^i$ as a section of $T^*X \otimes \phi^*(TM)$, and
$B_{1i}$ as a section of $T^*X \otimes \phi^*(T^*M)$.
We further introduce
$B_{2i}$ as a section of $\wedge^2 T^*X \otimes \phi^*(T^* M)$.


We propose the following action induced from a generalized complex structure:
\begin{eqnarray}
&& S = S_0 +S_1;
\nonumber \\
&& 
S_0 =  \int_{X} - B_{2i} d \phi^i + B_{1i} d A{}^i,
\nonumber \\
&& S_1 
= \int_{X} - J^i{}_j B_{2i} A{}^j
- P^{ij} B_{2i} B_{1j} 
+ \frac{1}{2} 
\frac{\partial Q_{jk}}{\partial \phi^{i}}
A{}^i A{}^j A^k
\nonumber \\ 
&& \quad \quad \ 
+  \frac{1}{2} \left(
- \frac{\partial J^k{}_{j}}{\partial \phi^i} 
+ \frac{\partial J^k{}_{i}}{\partial \phi^j} \right)
A{}^i A{}^j B_{1k}
+  \frac{1}{2} 
\frac{\partial P^{jk}}{\partial \phi^i} 
A{}^i B_{1j} B_{1k}.
\label{gcssigma}
\end{eqnarray}
(\ref{gcssigma}) is represented as
\begin{eqnarray}
&& S_0 = \int_X - \langle 0 + B_{2}, d ( \phi + 0) \rangle
+ \frac{1}{2} \langle A + B_1, d (A + B_1) \rangle
+ \mbox{total derivative},
\nonumber \\
&& S_1 = \int_X  - \langle 0 + B_{2}, \calJ  (A + B_{1}) \rangle
- \frac{1}{2} \langle  A + B_1, A^i \frac{\partial \calJ}{\partial \phi^i}
(A + B_1) \rangle.
\end{eqnarray}
We can confirm that 
if $J$, $P$ and $Q$ are components of 
a generalized complex structure (\ref{gcstcomp}),
the action (\ref{gcssigma})
is invariant under the following gauge transformation 
\begin{eqnarray}
\brs \phi^i &=& J^i{}_j c^j + P^{ij} t_{1j},
\nonumber \\
\brs A^i &=& d c^i - P^{ji} t_{2j}
+ \left( 
- \frac{\partial J^i{}_k}{\partial \phi^j}
+ \frac{\partial J^i{}_j}{\partial \phi^k} \right) A^j c^k
+ \frac{\partial P^{ki}}{\partial \phi^j} 
(A^j t_{1k} - c^j B_{1k}),
\nonumber \\
\brs B_{1i} &=& d t_{1i} - J^j{}_i t_{2j}
+ \left( 
\frac{\partial Q_{jk}}{\partial \phi^{i}}
+ \frac{\partial Q_{ki}}{\partial \phi^{j}}
+ \frac{\partial Q_{ij}}{\partial \phi^{k}} \right) A^j c^k
\nonumber \\
&& 
+ \left( 
- \frac{\partial J^k{}_j}{\partial \phi^i}
+ \frac{\partial J^k{}_i}{\partial \phi^j} \right) (A^j t_{1k} - c^j B_{1k})
+ \frac{\partial P^{jk}}{\partial \phi^i} B_{1j} t_{1k},
\nonumber \\
\brs B_{2i} &=& d t_{2i} 
- \frac{\partial J^j{}_k}{\partial \phi^i} (B_{2j} c^k + t_{2j} A^k)
- \frac{\partial P^{jk}}{\partial \phi^i} (B_{2j} t_{1k} + t_{2j} B_{1k})
\nonumber \\
&& 
+ \frac{\partial}{\partial \phi^i} 
\left( 
\frac{\partial Q_{kl}}{\partial \phi^{j}} 
+ \frac{\partial Q_{lj}}{\partial \phi^{k}} 
+ \frac{\partial Q_{jk}}{\partial \phi^{l}} 
\right) A^j A^k c^l
+ \frac{1}{2} \frac{\partial}{\partial \phi^i} \left( 
- \frac{\partial J^m{}_k}{\partial \phi^j}
+ \frac{\partial J^m{}_j}{\partial \phi^k} \right) 
(A^j A^k t_{1m} + 2 A^j c^k B_{1m})
\nonumber \\
&& 
+ \frac{1}{2} \frac{\partial^2 P^{kl}}{\partial \phi^i \partial \phi^j} 
(2 A^j B_{1k} t_{1l} + c^j B_{1k} B_{1l}),
\end{eqnarray}
where $c^i$ and $t_{1i}$ are $0$-form gauge parameters,
and $t_{2i}$ is a $1$-form gauge parameter.
We call this model the three dimensional 
generalized complex sigma model.
Precisely speaking, 
the action (\ref{gcssigma})
is gauge invariant if and only if the condition 
\begin{eqnarray}
&& \calA{}^{ijk} = \calB{}_i{}^{jk} = \calC{}_{ij}{}^k = 0,
\nonumber \\
&& \frac{\partial\calD{}_{ijk}}{\partial\phi^{l}} 
+ (ijkl \;\; \rm{cyclic}) = 0,
\label{cond}
\end{eqnarray}
is satisfied.

We can extend the model in case of a twisted generalized 
complex structure $H \neq 0$.
We define the following action as a three dimensional 
topological field theory induced from a twisted generalized complex 
structure: 
\begin{eqnarray}
&& S_H = S_{H0} +S_{H1};
\nonumber \\
&& 
S_{H0} =  \int_{X} - B_{2i} d \phi^i + B_{1i} d A{}^i,
\nonumber \\
&& S_{H1} 
= \int_{X} - J^i{}_j B_{2i} A{}^j
- P^{ij} B_{2i} B_{1j} 
+ \frac{1}{2} 
\left( J^l{}_{i} H_{jkl} + \frac{\partial Q_{jk}}{\partial \phi^{i}}
\right) A{}^i A{}^j A^k
\nonumber \\ 
&& \quad \quad \ 
+  \frac{1}{2} \left(- P^{kl} H_{ijl} 
- \frac{\partial J^k{}_{j}}{\partial \phi^i} 
+ \frac{\partial J^k{}_{i}}{\partial \phi^j} \right)
A{}^i A{}^j B_{1k}
+  \frac{1}{2} 
\frac{\partial P^{jk}}{\partial \phi^i} 
A{}^i B_{1j} B_{1k}.
\label{tgcssigma}
\end{eqnarray}
We call this model a three dimensional twisted generalized complex sigma model.
(\ref{tgcssigma}) is written as
\begin{eqnarray}
&& S_0 = \int_X - \langle 0 + B_{2}, d ( \phi + 0) \rangle
+ \frac{1}{2} \langle A + B_1, d (A + B_1) \rangle
+ \mbox{total derivative},
\nonumber \\
&& S_1 = \int_X - \langle 0 + B_{2} , \calJ  (A + B_{1}) \rangle
\nonumber \\
&& \qquad
- \frac{1}{2} \langle  A + B_1 , 
\left[ \left( 
\begin{array}{cc}
H_{ijk} A^k & 0 \\
0 & 0 \\
\end{array}
\right) \calJ
+ A^i \frac{\partial \calJ}{\partial \phi^i}
\right] (A + B_1) \rangle.
\end{eqnarray}
If and only if the condition 
\begin{eqnarray}
&& \calA_H{}^{ijk} = \calB_H{}_i{}^{jk} = \calC_H{}_{ij}{}^k = 0,
\nonumber \\
&& \frac{\partial\calD_H{}_{ijk}}{\partial\phi^{l}} 
+ (ijkl \;\; \rm{cyclic}) = 0,
\label{condh}
\end{eqnarray}
is satisfied,  
the action (\ref{tgcssigma}) is invariant under the gauge transformation 
\begin{eqnarray}
\brs \phi^i &=& J^i{}_j c^j + P^{ij} t_{1j},
\nonumber \\
\brs A^i &=& d c^i - P^{ji} t_{2j}
+ \left( - P^{il} H_{jkl} 
- \frac{\partial J^i{}_k}{\partial \phi^j}
+ \frac{\partial J^i{}_j}{\partial \phi^k} \right) A^j c^k
+ \frac{\partial P^{ki}}{\partial \phi^j} 
(A^j t_{1k} - c^j B_{1k}),
\nonumber \\
\brs B_{1i} &=& d t_{1i} - J^j{}_i t_{2j}
+ \left( J^l{}_{i} H_{jkl} 
+ J^l{}_{j} H_{kil} 
+ J^l{}_{k} H_{ijl} 
+ \frac{\partial Q_{jk}}{\partial \phi^{i}}
+ \frac{\partial Q_{ki}}{\partial \phi^{j}}
+ \frac{\partial Q_{ij}}{\partial \phi^{k}} \right) A^j c^k
\nonumber \\
&& 
+ \left( - P^{kl} H_{ijl} 
- \frac{\partial J^k{}_j}{\partial \phi^i}
+ \frac{\partial J^k{}_i}{\partial \phi^j} \right) (A^j t_{1k} - c^j B_{1k})
+ \frac{\partial P^{jk}}{\partial \phi^i} B_{1j} t_{1k},
\nonumber \\
\brs B_{2i} &=& d t_{2i} 
- \frac{\partial J^j{}_k}{\partial \phi^i} (B_{2j} c^k + t_{2j} A^k)
- \frac{\partial P^{jk}}{\partial \phi^i} (B_{2j} t_{1k} + t_{2j} B_{1k})
\nonumber \\
&& 
+ \frac{\partial}{\partial \phi^i} 
\left( J^m{}_{j} H_{klm} + J^m{}_{k} H_{ljm} + J^m{}_{l} H_{jkm} 
+ \frac{\partial Q_{kl}}{\partial \phi^{j}} 
+ \frac{\partial Q_{lj}}{\partial \phi^{k}} 
+ \frac{\partial Q_{jk}}{\partial \phi^{l}} 
\right) A^j A^k c^l
\nonumber \\
&& 
+ \frac{1}{2} \frac{\partial}{\partial \phi^i} \left( - P^{lm} H_{jkm} 
- \frac{\partial J^m{}_k}{\partial \phi^j}
+ \frac{\partial J^m{}_j}{\partial \phi^k} \right) 
(A^j A^k t_{1m} + 2 A^j c^k B_{1m})
\nonumber \\
&& 
+ \frac{1}{2} \frac{\partial^2 P^{kl}}{\partial \phi^i \partial \phi^j} 
(2 A^j B_{1k} t_{1l} + c^j B_{1k} B_{1l}),
\label{agauge}
\end{eqnarray}
We can obtain (\ref{gcssigma}) as the $H = 0$ theory in (\ref{tgcssigma}).
Moreover the action (\ref{tgcssigma}) (and (\ref{gcssigma}))
is invariant under the $b$-transformation
(\ref{btrans}) and (\ref{btransj}) by simple calculation
if we define the $b$-transformation on the fields 
\begin{eqnarray}
&& \hat{\phi}^i = \phi^i,
\nonumber \\
&& \hat{A}^i = A^i,
\nonumber \\
&& \hat{B}_{1i} = B_{1i} + b_{ij} A^j,
\nonumber \\
&& \hat{B}_{2i} =  B_{2i} 
- \frac{1}{2} \frac{\partial b_{jk}}{\partial \phi^i} A^j A^k.
\label{btransfield}
\end{eqnarray}

The condition which the action is gauge invariant is not 
$\calD{}_{ijk} = 0$ but 
in (\ref{cond}), and
$\calD_H{}_{ijk} = 0$ 
but $\partial\calD_H{}_{ijk}/\partial\phi^{l}
+ (ijkl \; \rm{cyclic}) = 0$
in (\ref{condh}).
This is because 
our model is $b$-transformation invariant and
$H{}_{ijk}$ has $b$-transformation ambiguity
(\ref{btransj}), that is,
$H$ is defined as a cohomology class in $H^3(M)$
in a (twisted) generalized complex structure.

Let us consider mirror transformation in our model.
Mirror transform is exchange of 
the complex structure 
$J^i{}_j$
and the symplectic structure
$P^{ij}$ and $Q_{ij}$.

A simplest mirror transformation $m_0$ is exchange
\begin{eqnarray}
J^i{}_j A{}^j \leftrightarrow P^{ij} B_{1j}, 
\nonumber \\
- J^j{}_i  B_{1j} \leftrightarrow Q_{ij} A^{j}.
\end{eqnarray}
The action (\ref{tgcssigma}) is invariant under this symmetry.

A general mirror transform $m: TM \oplus T^*M \rightarrow
TM \oplus T^*M$ is a $Z_2$ transformation with
\begin{eqnarray}
&& m^2 = 1
\nonumber \\
&& m(J^i{}_j A{}^j + P^{ij} B_{1j}) = J^i{}_j A{}^j + P^{ij} B_{1j}, 
\nonumber \\
&& m(- J^j{}_i  B_{1j} + Q_{ij} A^{j}) = - J^j{}_i  B_{1j} + Q_{ij} A^{j}.
\label{mirror}
\end{eqnarray}
We can easily confirm that the action (\ref{tgcssigma}) is invariant
under $m$.

\section{Derivation from 3D Nonlinear Gauge Theory}
\noindent
In the paper \cite{Ikeda:2000yq}, We have proposed 
3D nonlinear BF theory as a general topological field theory 
of Schwarz type in three dimension.
A 3D (twisted) generalized complex sigma model is 
a special model of a 3D nonlinear BF theory.

Let $X$ be a three-dimensional manifold
and $M$ be a target manifold
of a smooth map $\phi:X \to M$  with local
coordinate expression $\{\phi^i\}$.
We also have a vector bundle $E$ over $M$
with $A^a$ a section of $T^*X \otimes \phi^*(E)$.
%
We further introduce
$B_{1a}$ as a section of $T^*X \otimes \phi^*(E^*)$
and $B_{2i}$ as a section of $ \wedge^2 T^*X \otimes \phi^*(T^*M)$.
Hereafter, the letters $a, b, \cdots$ represent
indices on the fiber of $E$
and $i, j, \cdots$ represent indices on $M$ and $T^*M$.

3D nonlinear BF theory with nonlinear gauge symmetry 
has the following action 
\begin{eqnarray}
&& S = S_0 +S_1;
\nonumber \\
&& 
S_0 =  \int_{X} \left( - B_{2i} d \phi^i + B_{1a} d A{}^a 
\right),
\nonumber \\
&& S_1 
= \int_{X} ( f_{1a}{}^i (\phi) B_{2i} A{}^a 
+ f_{2}{}^{ib} (\phi) B_{2i} B_{1b} 
+  \frac{1}{6} f_{3abc} (\phi) A{}^a A{}^b A^c
\nonumber \\ 
&& \quad \quad \ 
+  \frac{1}{2} f_{4ab}{}^c (\phi) A{}^a A{}^b B_{1c}
+  \frac{1}{2} f_{5a}{}^{bc} (\phi) A{}^a B_{1b} B_{1c}
+  \frac{1}{6} f_6{}^{abc} (\phi) B_{1a} B_{1b} B_{1c}
),
\label{bf}
\end{eqnarray}
where 
the structure functions $f_1, \cdots, f_6$ satisfy the identities
\begin{eqnarray}
&& 
f_{1}{}_e{}^i f_{2}{}^{je} + f_{2}{}^{ie} f_{1}{}_e{}^j = 0, 
\nonumber \\
&& 
- \frac{\partial f_{1}{}_c{}^i}{\partial \phi^j} f_{1}{}_b{}^j
+ \frac{\partial f_{1}{}_b{}^i}{\partial \phi^j} f_1{}_c {}^j
+ f_1{}_e{}^i f_{4bc}{}^e + f_{2}{}^{ie} f_{3ebc} = 0, 
\nonumber \\
&&
 f_1{}_b{}^j \frac{\partial f_{2}{}^{ic}}{\partial \phi^j} 
- f_{2}{}^{jc} \frac{\partial f_1{}_b{}^i}{\partial \phi^j} 
+ f_1{}_e{}^i  f_{5b}{}^{ec} - f_{2}{}^{ie} f_{4eb}{}^c = 0, 
\nonumber \\
&&
- f_{2}{}^{jb} \frac{\partial f_{2}{}^{ic}}{\partial \phi^j} 
+ f_{2}{}^{jc} \frac{\partial f_{2}{}^{ib}}{\partial \phi^j} 
+ f_1{}_e{}^i f_{6}^{ebc} + f_{2}{}^{ie} f_{5e}{}^{bc} = 0, 
\nonumber \\
&&
- f_1{}_{[a}{}^j \frac{\partial f_{4bc]}{}^d}{\partial \phi^j} 
+ f_{2}{}^{jd} \frac{\partial f_{3abc}}{\partial \phi^j} 
+ f_{4e[a}{}^d f_{4bc]}{}^{e} + f_{3e[ab} f_{5c]}{}^{de} = 0, 
\nonumber \\
&&
- f_1{}_{[a}{}^j \frac{\partial f_{5b]}{}^{cd}}{\partial \phi^j} 
- f_{2}{}^{j[c} \frac{\partial f_{4ab}{}^{d]}}{\partial \phi^j} 
+ f_{3eab} f_6{}^{ecd} 
+ f_{4e[a}{}^{[d} f_{5b]}{}^{c]e} + f_{4ab}{}^e f_{5e}{}^{cd} = 0, 
\nonumber \\
&&
- f_1{}_a{}^j \frac{\partial f_{6}{}^{bcd}}{\partial \phi^j} 
+ f_{2}{}^{j[b} \frac{\partial f_{5a}{}^{cd]}}{\partial \phi^j} 
+ f_{4ea}{}^{[b} f_6{}^{cd]e} + f_{5e}{}^{[bc} f_{5a}{}^{d]e} = 0, 
\nonumber \\
&&
- f_{2}{}^{j[a} \frac{\partial f_{6}{}^{bcd]}}{\partial \phi^j} 
+ f_6{}^{e[ab} f_{5e}{}^{cd]} = 0, 
\nonumber \\
&&
- f_1{}_{[a}{}^j \frac{\partial f_{3bcd]}}{\partial \phi^j} 
+ f_{4[ab}{}^{e} f_{3cd]e} = 0.
\label{3dJacobi2}
\end{eqnarray}

This theory has a Courant algebroid structure.
A Courant algebroid \cite{Courant}
is a vector bundle $E \rightarrow M$
with a nondegenerate symmetric bilinear form
$\bracket{\cdot}{\cdot}$ 
on the bundle, a bilinear operation (a Dorfman bracket) 
$\circ$ on $\Gamma(E)$ (the
space of sections on $E$), and a bundle map (called the anchor) 
$\rho: E \rightarrow TM$ satisfying the following properties:
\begin{eqnarray}
&& 1) \quad e_1 \circ (e_2 \circ e_3) = (e_1 \circ e_2) \circ e_3 
+ e_2 \circ (e_1 \circ e_3), \nonumber \\
&& 2) \quad \rho(e_1 \circ e_2) = [\rho(e_1), \rho(e_2)], \nonumber \\
&& 3) \quad e_1 \circ F e_2 = F (e_1 \circ e_2)
+ (\rho(e_1)F)e_2, \nonumber \\
&& 4) \quad e_1 \circ e_2 = \frac{1}{2} {\cal D} \bracket{e_1}{e_2},
\nonumber \\ 
&& 5) \quad \rho(e_1) \bracket{e_2}{e_3}
= \bracket{e_1 \circ e_2}{e_3} + \bracket{e_2}{e_1 \circ e_3},
  \label{courantdef}
\end{eqnarray}
where 
$e_1, e_2$, and $e_3$ are sections of $E$;
$F$ is a function on $M$;
${\cal D}$ is a map from functions on $M$ to $\Gamma(E)$ 
and is defined by
$\bracket{{\cal D}F}{e} = \rho(e) F$.

If we take a local basis, 
Eq.(\ref{3dJacobi2}) is equivalent to the relations of 
structure functions of a Courant algebroid on a vector bundle 
$E \oplus E^*$ over $M$:
Symmetric bilinear form $\bracket{\cdot}{\cdot}$ 
is defined from the natural pairing of $E$ and $E^*$.
That is,
$\bracket{e_a}{e_b} = \bracket{e^a}{e^b} = 0$ and 
$\bracket{e_a}{e^b} = \delta_a{}^b$
if $\{e_a\}$ is a basis of sections of $E^*$
and $\{e^a\}$ is that of $E$.
The bilinear form $\circ$ and the anchor $\rho$ are represented 
as follows:
\begin{eqnarray}
&& e{}^a \circ e{}^b = - f_{5c}{}^{ab}(\phi) e^c
- f_{6}{}^{abc}(\phi) e_{c}, \nonumber \\
&& e^a \circ e_{b} = - f_{4bc}{}^{a}(\phi) e^c
+ f_{5b}{}^{ac}(\phi) e_{c}, \nonumber \\
&& e_{a} \circ e_{b} = - f_{3abc}(\phi) e^c
- f_{4ab}{}^{c}(\phi) e_{c}, \nonumber \\
&& \rho(e^a) = - f_{2}{}^{ia}(\phi) 
\frac{\partial}{\partial \phi^i}, \nonumber \\
&& \rho(e_a) = - f_{1a}{}^i(\phi) 
\frac{\partial}{\partial \phi^i}.
\end{eqnarray}

If we take $E = TM$ and $E^* = T^*M$, 
and we set
\begin{eqnarray}
&& 
f_1{}_{j}{}^i = - J^{i}{}_j,
\nonumber \\ 
&& 
f_2{}^{ij} = - P^{ij}, 
\nonumber \\ 
&& 
f_{3ijk} = J^l{}_{i} H_{jkl} + J^l{}_{j} H_{kil} + J^l{}_{k} H_{ijl} 
+ \frac{\partial Q_{jk}}{\partial \phi^{i}}
+ \frac{\partial Q_{ki}}{\partial \phi^{j}}
+ \frac{\partial Q_{ij}}{\partial \phi^{k}},
\nonumber \\ 
&& 
f_{4ij}{}^{k} = - P^{kl} H_{ijl} 
- \frac{\partial J^k{}_{j}}{\partial \phi^{i}}
+ \frac{\partial J^k{}_{i}}{\partial \phi^{j}},
\nonumber \\ 
&& 
f_{5i}{}^{jk} = \frac{\partial P^{jk}}{\partial \phi^{i}},
\nonumber \\ 
&& 
f_6{}^{ijk} = 0,
\label{f1-f6}
\end{eqnarray}
we can find that 3D nonlinear gauge theory (\ref{bf}) reduces to the
(twisted) generalized complex sigma model (\ref{tgcssigma}), and
(\ref{3dJacobi2}) reduces to (\ref{condh}).

\section{Zucchini Model as a Boundary Action}
Zucchini has proposed a topological sigma model 
with a generalized complex structure on 
two dimensional worldsheet \cite{Zucchini:2004ta}.
He has called the model the Hitchin sigma model.

First we assume $H = 0$.
The Zucchini' s action is 
\begin{eqnarray}
S_Z = \int_{\Sigma}
B_{1i} d \phi^i 
+ \frac{1}{2} P^{ij} B_{1i} B_{1j} + \frac{1}{2} Q_{ij} d \phi^i d \phi^j 
+ J^i{}_j B_{1i} d \phi^j,
\label{zucchini}
\end{eqnarray}
where $\Sigma$ is a two dimensional world sheet and 
$\phi^i$ and $B_{1i}$ are restricted on $\Sigma$.
We consider this model as a boundary theory. 
That is, we consider a three dimensional world volume $X$ such that 
$\Sigma = \partial X$,
and we regard the action (\ref{zucchini}) as 
\begin{eqnarray}
S_Z &=& \int_{X}
d \left( B_{1i} d \phi^i 
+ \frac{1}{2} P^{ij} B_{1i} B_{1j} + \frac{1}{2} Q_{ij} d \phi^i d \phi^j 
+ J^i{}_j B_{1i} d \phi^j \right)
\nonumber \\
&=& \int_{X}
d B_{1i} d \phi^i 
+ \frac{1}{2} \frac{\partial P^{ij}}{\partial \phi^k} d \phi^k B_{1i} B_{1j} 
+ P^{ij} d B_{1i} B_{1j} 
+ \frac{1}{2} \frac{\partial Q_{ij}}{\partial \phi^k} 
d \phi^k d \phi^i d \phi^j 
\nonumber \\
&&
+ \frac{\partial J^i{}_j}{\partial \phi^k} d \phi^k B_{1i} d \phi^j 
+ J^i{}_j d B_{1i} d \phi^j.
\end{eqnarray}
We introduce a $1$-form $A^i$ and a $2$-from $B_{2i}$ such that
$A^i = d \phi^i$ and $B_{2i} = - d B_{1i}$.
If we introduce two Lagrange multiplier fields, a $2$-form $Y_{2i}$ 
and a $1$-form $Z^i$, we obtain an equivalent action 
\begin{eqnarray}
S_Z &=& \int_{X}
- B_{2i} A^i 
+ \frac{1}{2} \frac{\partial P^{ij}}{\partial \phi^k} A^k B_{1i} B_{1j} 
- P^{ij} B_{2i} B_{1j} 
+ \frac{1}{2} \frac{\partial Q_{jk}}{\partial \phi^i} A^i A^j A^k 
\nonumber \\
&&
+ \frac{\partial J^i{}_j}{\partial \phi^k} A^k B_{1i} A^j 
- J^i{}_j B_{2i} A^j
\nonumber \\
&&
+ (A^i - d \phi^i ) Y_{2i}
+ (B_{2i} + d B_{1i} ) Z^i.
\label{baction}
\end{eqnarray}
Moreover we redefine $Y_{2i}$ and $Z^i$ as 
$Y^\prime_{2i} = Y_{2i} - B_{2i}$
and $Z^{\prime i} = Z^i - A^i$.
Then (\ref{baction}) is 
\begin{eqnarray}
&& S_Z = S_a + S_{Zb};
\nonumber \\
&& S_{a} =  
\int_{X} - Y^\prime_{2i} d \phi^i + d B_{1i} Z^{\prime i}
+ Y^\prime_{2i} A^i + B_{2i} Z^{\prime i}
+ B_{2i} A^i,
\nonumber \\
&& S_{Zb}
= S
\nonumber \\
&& \qquad 
= \int_{X} - B_{2i} d \phi^i + d B_{1i} A{}^i 
- J^i{}_j B_{2i} A{}^j
- P^{ij} B_{2i} B_{1j} 
+ \frac{1}{2} 
\frac{\partial Q_{jk}}{\partial \phi^{i}}
A{}^i A{}^j A^k
\nonumber \\ 
&& \quad \quad \ 
+  \frac{1}{2} \left(
- \frac{\partial J^k{}_{j}}{\partial \phi^i} 
+ \frac{\partial J^k{}_{i}}{\partial \phi^j} \right)
A{}^i A{}^j B_{1k}
+  \frac{1}{2} 
\frac{\partial P^{jk}}{\partial \phi^i} 
A{}^i B_{1j} B_{1k}.
\label{boundarysigma}
\end{eqnarray}
$S_{Zb}$ coincides with (a partial derivative of) our 
generalized complex sigma model (\ref{gcssigma}) 
and $S_a$ are terms independent of a generalized complex structure.
Therefore $S_{Zb}$ and $S_{Z}$ are constructed from the same generalized 
complex structure.
We have found that three dimensional topological field theory
equivalent to the Zucchini's  model, which is derived from the same 
generalized complex structure.

We make similar discussion 
for a twisted generalized complex structure with $H \neq 0$.
From (\ref{tgcssigma}), it is natural to consider the action 
\begin{eqnarray}
&& S_T = S_{a} + S_{Tb};
\nonumber \\
&& S_a =  
\int_{X} - Y^\prime_{2i} d \phi^i + d B_{1i} Z^{\prime i}
+ Y^\prime_{2i} A^i + B_{2i} Z^{\prime i}
+ B_{2i} A^i,
\nonumber \\
&&
S_{Tb} = S_{H} 
\nonumber \\ 
&& \quad \ \
=  \int_{X} - B_{2i} d \phi^i + B_{1i} d A{}^i 
- J^i{}_j B_{2i} A{}^j
- P^{ij} B_{2i} B_{1j} 
+ \frac{1}{2} 
\left( J^l{}_{i} H_{jkl} + \frac{\partial Q_{jk}}{\partial \phi^{i}}
\right) A{}^i A{}^j A^k
\nonumber \\ 
&& \quad \quad \ 
+  \frac{1}{2} \left(- P^{kl} H_{ijl} 
- \frac{\partial J^k{}_{j}}{\partial \phi^i} 
+ \frac{\partial J^k{}_{i}}{\partial \phi^j} \right)
A{}^i A{}^j B_{1k}
+  \frac{1}{2} 
\frac{\partial P^{jk}}{\partial \phi^i} 
A{}^i B_{1j} B_{1k}.
\label{tboundarysigma}
\end{eqnarray}
We rewrite the action (\ref{tboundarysigma}) 
by the procedure with the modification from (\ref{boundarysigma}) to 
(\ref{zucchini}).
If we remove $Y_{2i}$, $Z^{i}$, $A^i$ and $B_{2i}$
by means of the equations of motion, we obtain
\begin{eqnarray}
S_{T} &=& \int_{\Sigma}
B_{1i} d \phi^i 
+ \frac{1}{2} P^{ij} B_{1i} B_{1j} + \frac{1}{2} Q_{ij} d \phi^i d \phi^j 
+ J^i{}_j B_{1i} d \phi^j
\nonumber \\
&&
+ \frac{1}{2} \int_{X}
J^l{}_i H_{jkl} d \phi^i d \phi^j d \phi^k
- P^{kl} H_{ijl} B_{1k} d \phi^i d \phi^j,
\label{tzucchini}
\end{eqnarray}
the second line in (\ref{tzucchini}) is 'Wess-Zumino' terms,
which is integration on $X$.
However these 'Wess-Zumino' terms are different from Zucchini's $H$-term.
If we consider the following 3D action as the beginning
\begin{eqnarray}
&& S_{ZH} = S_a + S_{ZHb};
\nonumber \\
&& S_a =  
\int_{X} - Y^\prime_{2i} d \phi^i + d B_{1i} Z^{\prime i}
+ Y^\prime_{2i} A^i + B_{2i} Z^{\prime i}
+ B_{2i} A^i,
\nonumber \\
&&
S_{ZHb} =  \int_{X} - B_{2i} d \phi^i + B_{1i} d A{}^i 
- J^i{}_j B_{2i} A{}^j
- P^{ij} B_{2i} B_{1j} 
+ \frac{1}{2} 
\left( H_{ijk} + \frac{\partial Q_{jk}}{\partial \phi^{i}}
\right) A{}^i A{}^j A^k
\nonumber \\ 
&& \quad \quad \ 
+  \frac{1}{2} \left(
- \frac{\partial J^k{}_{j}}{\partial \phi^i} 
+ \frac{\partial J^k{}_{i}}{\partial \phi^j} \right)
A{}^i A{}^j B_{1k}
+  \frac{1}{2} 
\frac{\partial P^{jk}}{\partial \phi^i} 
A{}^i B_{1j} B_{1k},
\label{ztboundarysigma}
\end{eqnarray}
we obtain a boundary action
\begin{eqnarray}
S_{ZH} &=& \int_{\Sigma}
B_{1i} d \phi^i 
+ \frac{1}{2} P^{ij} B_{1i} B_{1j} + \frac{1}{2} Q_{ij} d \phi^i d \phi^j 
+ J^i{}_j B_{1i} d \phi^j
\nonumber \\
&&
+ \frac{1}{2} \int_{X}
H_{ijk} d \phi^i d \phi^j d \phi^k,
\label{ztzucchini}
\end{eqnarray}
which coincides with Zucchini's proposal.

The point is $b$-transformation property of the action 
(\ref{zucchini}).
The Zucchini's action (\ref{zucchini}) is not invariant under 
the $b$-transformation
\begin{eqnarray}
\hat{S}_Z = S_Z - \int_{\Sigma} b_{ij} d \phi^i d \phi^j,
\label{btzucchini}
\end{eqnarray}
where the $b$-transformation is defined as
\begin{eqnarray}
&& \hat{\phi}^{i} = \phi^i, 
\nonumber \\
&& \hat{B}_{1i} = B_{1i} + b_{ij} d \phi^j.
\end{eqnarray}
The equivalent 3D action (\ref{boundarysigma}) also 
changes under the $b$-transformation
\begin{eqnarray}
\hat{S}_Z = S_Z - \frac{1}{2} \int_{X} 
\frac{\partial b_{jk}}{\partial \phi^i} A^i A^j A^k
- \frac{1}{2} \int_{X} d ( b_{ij} A^i A^j),
\label{btransbtzucchini}
\end{eqnarray}
where $b$-transformation is defined by (\ref{btrans}), 
(\ref{btransfield}) and
\begin{eqnarray}
&& \hat{Y}^{\prime}_{2i} = 
\frac{1}{2} \frac{\partial b_{jk}}{\partial \phi^i} A^j A^k 
+ b_{ij} d A^j 
- \frac{\partial b_{jk}}{\partial \phi^i} A^j d \phi^k 
- J^l{}_k \frac{\partial b_{jl}}{\partial \phi^i} A^j A^k 
- P^{lk} \frac{\partial b_{jl}}{\partial \phi^i} A^j B_{1k}, 
\nonumber \\
&& \hat{Z}^{\prime i} = Z^{\prime i},
\end{eqnarray}
however in (\ref{btransfield}),
the $b$-transformation of $B_{2i}$ is changed to
\begin{eqnarray}
\hat{B}_{2i} = {B}_{2i} - d ( b_{ij} A^j),
\end{eqnarray}
because of consistency with equations of motion.

Since $H$ is closed, we can {\it locally} write $H$ 
by a $2$-form $q$ 
on $M$ as
\begin{eqnarray}
H_{ijk} = \frac{1}{3} 
\left( \frac{\partial q_{jk}}{\partial \phi^i} 
+ \frac{\partial q_{ki}}{\partial \phi^j} 
+ \frac{\partial q_{ij}}{\partial \phi^k} \right).
\end{eqnarray}
Since there exists the
$\frac{\partial b_{jk}}{\partial \phi^i} A^i A^j A^k$ term in
(\ref{btransbtzucchini}),
the $H$-term can be absorbed to $Q$ by a 
{\it local} $b$-transformation $q_{ij} = b_{ij}$ 
in the action (\ref{ztboundarysigma}), and
we obtain (\ref{boundarysigma}).
In other words, the $H$ term in (\ref{ztboundarysigma}) is consistent 
up to $H$-exact terms as a global theory, 
the theory is meaningful only as a cohomology class in $H^3(M)$.
It describes gerbe gauge transformation dependence 
\cite{Zucchini:2004ta}.

If we set $Q_{ij} = J^i{}_j = 0$ in  (\ref{ztzucchini}),
we obtain the WZ-Poisson sigma model
\begin{eqnarray}
S_{WZP} &=& \int_{\Sigma}
B_{1i} d \phi^i 
+ \frac{1}{2} P^{ij} B_{1i} B_{1j}
+ \frac{1}{2} \int_{X}
H_{jkl} d \phi^i d \phi^j d \phi^k,
\label{WZpoisson}
\end{eqnarray}
and from (\ref{ztboundarysigma}), we obtain 
an equivalent 3D sigma model \begin{eqnarray}
&& S_{WZP} = S_a + S_{WZPb};
\nonumber \\
&& S_a =  
\int_{X} - Y^\prime_{2i} d \phi^i + d B_{1i} Z^{\prime i}
+ Y^\prime_{2i} A^i + B_{2i} Z^{\prime i}
+ B_{2i} A^i,
\nonumber \\
&& S_{WZPb}
= \int_{X} - B_{2i} d \phi^i + d B_{1i} A{}^i 
- P^{ij} B_{2i} B_{1j} 
+ \frac{1}{2} 
H_{ijk}
A{}^i A{}^j A^k
\nonumber \\ 
&& \quad \quad \ 
+  \frac{1}{2} 
\frac{\partial P^{jk}}{\partial \phi^i} 
A{}^i B_{1j} B_{1k},
\label{boundaryWZpoisson}
\end{eqnarray}

\section{Conclusions and Discussion}
\noindent
We have constructed a three-dimensional topological sigma model based 
on a (twisted) generalized complex structure on a target generalized 
complex manifold.
The theory is invariant under the diffeomorphism on the world volume, 
the $b$-transformation on the (twisted) generalized complex structure and
the mirror symmetries in a generalized complex geometry.

We derive the Zucchini's two dimensional topological sigma model with 
a generalized complex structure as a boundary action of our three dimensional 
action plus terms independent of a generalized complex structure.
In case of a twisted generalized complex structure, 
the $H$ term in the Zucchini action is different from 
our proposal, however 
the 'Wess-Zumino' terms in the Zucchini action 
can be generated {\it locally} by the $b$-transformation 
because 
Zucchini action has a gerbe gauge transformation dependence.

This model will be useful to investigate mirror symmetry.
We have analyzed only a classical theory in this paper.
It is important to investigate a quantum theory.
There exist several topological and non topological string theories
with a generalized complex structure other than the Zucchini's model.
Relations to other known theories based on a generalized 
complex structure will be interesting.

\section*{Acknowledgements}

We would like to thank K.~-I.~Izawa for valuable comments and 
discussions.

\newcommand{\bibit}{\sl}



\vfill\eject
\end{document}